\documentstyle[12pt]{article}

\author{M. Temple-Raston\\
Department of Mathematics,\\
Concordia University,\\
1455 de Maisonneuve Blvd. West,\\
Montreal, Quebec H3G 1M8 CANADA}
\title{Electric monopoles and the Montonen-Olive conjecture
}

\begin{document}

\maketitle
\begin{abstract}
We introduce a topological field theory with a Bogomol'nyi structure
permitting BPS electric, magnetic and dyonic monopoles. From the general
arguments given by Montonen and Olive the particle spectrum and mass compare
favourably with that of the intermediate vector bosons. In most, if not in
all, of its essential features the topological field theory introduced here
provides an example of a dual field theory, the existence of which was
conjectured by Montonen and Olive.
\end{abstract}

In \cite{olive} Montonen and Olive conjectured the existence of a Lagrangian
field theory `dual' to Georgi-Glashow theory that would possess electric
monopole solutions analogous to the BPS magnetic monopole. In favour of this
conjecture Montonen and Olive argued that classical properties of the BPS
monopole in the alleged dual field theory, namely the particle spectrum, the
mass, and the long-range intermonopole force, correspond to the anticipated
quantum properties of heavy gauge bosons (with a massless Higgs particle).
The quantum corrections to the classical BPS monopole are unknown, but
Montonen and Olive suggested that the corrections might be small, or, vanish
entirely, due to symmetry. Lending some support to this expectation are
recent studies of interacting classical BPS\ magnetic monopoles, showing
that a classical solitonic theory of BPS monopoles is capable of capturing
dynamical behaviour usually associated with quantum mechanics \cite{tra}.

Sixteen years after the conjecture was first made and despite the compelling
evidence, not one explicit example of a dual field theory is known. This
suggests that the dual field theory, if it exists, is structurally rather
different from any theory previously considered. In \cite{topdyon} we argued
that a topological field theory of a form first proposed by Horowitz \cite
{horowitz} might lead to electric monopoles, and, perhaps, a dual field
theory. The topological field theory was generalised in \cite{topinst} to
provide the Lagrangian with a Bogomol'nyi structure similar to the
Bogomol'nyi structure leading to instantons in Yang-Mills theory. By virtue
of the many similarities with self-dual instantons, solutions to the
Bogomol'nyi equations in \cite{topinst} are called `topological instantons'.
It is well-known that self-dual instantons and BPS\ magnetic monopoles are
closely related by dimensional reduction with a gauge symmetry. By a similar
process of dimensional reduction on topological instantons, a theory of
topological monopoles is developed \cite{topsol}. In this paper we study the
theory of topological monopoles to see whether it can be viewed as a dual
field theory.

Let $G$ be a compact, semi-simple Lie group and $P$ a principal $G$-bundle
over a three-manifold $M_3$. Denote the space of connections on $P$ by $%
{\cal A}(P)$. We represent by $E$ the vector bundle over $M_3$ associated to
$P$ by the adjoint representation. Associated with each connection $A\in
{\cal A}(P)$ is an exterior covariant derivative $D^A$ acting on sections of
$E$. The covariant derivative defines a curvature $H^A$ for the vector
bundle $E$ by $D^AD^As=sH^A$, where $s$ is a section of the vector bundle $%
E\rightarrow M$. The curvature $H^A$ can be interpreted as a 2-form on $M$
taking values in $E$. We also introduce an equivariant Lie algebra valued
Higgs field, $\Phi _A$ , on $M_3$ associated with the connection $A$. Our
theory is defined by the energy functional ${\cal E}(A,B,\Phi _A,\Phi _B)$
given by
\begin{equation}
\label{TFT3}
\begin{array}{ccc}
\int_{M_3}<(I_E\otimes K^B)\wedge (I_E\otimes D^B\Phi _B)> & - &
<(I_E\otimes K^B)\wedge (D^A\Phi _A\otimes I_E)> \\
-<(H^A\otimes I_E)\wedge (I_E\otimes D^B\Phi _B)> & + & (A\leftrightarrow
B,\Phi _A\leftrightarrow \Phi _B).
\end{array}
\end{equation}
$I_E$ is the identity transformation on the adjoint bundle, $E$. Note that
in (\ref{TFT3}) there are two curvatures $H^A,K^B$ and two Higgs fields $%
\Phi _A,\Phi _B$ corresponding to the two connections $A,B$. Finally, we
assume that there is an invariant positive-definite inner product $<\ >$ on $%
E_x$ which varies continuously with $x\in M$. \.The last term in (\ref{TFT3}%
) symmetrises the energy functional in the dependent fields. The energy
functional is rather complicated, but as we shall see it is constructed so
as to give certain geometric properties. The functional (\ref{TFT3}) can be
derived from a four-dimensional topological field theory using dimensional
reduction \cite{topsol}. The three-dimensional theory can be examined
independently of the four-dimensional theory, so a derivation of the
three-dimensional theory is not needed here. The variational field equations
arising from (\ref{TFT3}) are
\begin{equation}
\label{field3}
\begin{array}{c}
D^BH^A=0,\qquad \qquad D^BD^A\Phi _A=[H^A,\Phi _B], \\
D^AK^B=0,\qquad \qquad D^AD^B\Phi _B=[K^B,\Phi _A].
\end{array}
\end{equation}
The energy functional (\ref{TFT3}) can be rewritten as
\begin{equation}
\label{top3}
\begin{array}{ccc}
{\cal E}=\int_{M_3}<(H^A\otimes I_E-I_E\otimes K^B) & \wedge  & (D^A\Phi
_A\otimes I_E-I_E\otimes D^B\Phi _B)> \\
-\int_{M_3}<(D^A\Phi _A\otimes I_E)\wedge (H^A\otimes I_E)> & + &
(A\leftrightarrow B,\Phi _A\leftrightarrow \Phi _B).
\end{array}
\end{equation}
In this equation we begin to see the geometrical structure emerge out of the
energy functional (\ref{TFT3}). Let $E_A$ and $E_B$ be the vector bundle $E$
equipped with either the connection $A$ or $B$, respectively. Recall that
the curvature of the tensor product bundle $E_A\otimes E_B^{*}$ is given by $%
\Omega _{E_A\otimes E_B^{*}}=$$H^A\otimes I_E-I_E\otimes K^B$. This
curvature expression appears in (\ref{top3}). By defining $\Phi =\Phi
_A\otimes I_E-I_E\otimes \Phi _B$, then $D_{E_A\otimes E_B^{*}}\Phi =D^A\Phi
_A\otimes I_E-I_E\otimes D^B\Phi _B$, and the first integral in (\ref{top3})
is therefore a topological invariant. The Bogomol'nyi equations arising from
equation (\ref{TFT3}) are
\begin{equation}
\label{bog}
\begin{array}{c}
H^A\otimes I_E=I_E\otimes K^B \\
D^A\Phi _A\otimes I_E=I_E\otimes D^B\Phi _B
\end{array}
\end{equation}
The first equation in (\ref{bog}) is therefore a zero curvature condition on
the tensor product bundle $E_A\otimes E_B^{*}$. From an indices computation
we conclude that
\begin{equation}
\label{bog'}
\begin{array}{c}
H^A=K^B=iFI_E, \\
D^A\Phi _A=D^B\Phi _B=iEI_E.
\end{array}
\end{equation}
$F$ and $E$ are a real-valued two-form and one-form on $M_3$, respectively.
Solutions to (\ref{bog'}) automatically satisfy the variational field
equations (\ref{field3}). Moreover, solutions to the first equation are
projectively flat connections \cite{kob}. The projective flatness reduces
the gauge group to $U(n)$. For line bundles this carries no extra
information, but for bundles of rank greater than one projective flatness is
a strong condition. Projectively flat connections are closely related to
Einstein-Hermitian connections, which in turn express in differential
geometry the algebraic geometric concept of `stable vector bundle' \cite{kob}%
. As a result, projectively flat connections have a well-behaved moduli
space---like self-dual instantons and BPS magnetic monopoles. Unlike the
theory of BPS magnetic monopoles, however, the energy functional (\ref{top3}%
) is saturated at the Bogomol'nyi energy with either Bogomol'nyi equation in
(\ref{bog'}) satisfied. Of course the field configurations must still
satisfy the second-order variational field equations. The Bogomol'nyi energy
is given by
\begin{equation}
\label{energy}{\cal E}=-\int_{M_3}<(D^A\Phi _A\otimes I_E)\wedge (H^A\otimes
I_E)>-\int_{M_3}<(D^B\Phi _B\otimes I_E)\wedge (K^B\otimes I_E)>.
\end{equation}
The flexibility in obtaining the Bogomol'nyi energy, a feature not present
in the theory of magnetic monopoles, will lead to electric monopoles.

Since topological monopoles, if they exist, are analogous to the BPS
magnetic monopole field configurations, we shall use the same symmetry
breaking mechanism \cite{Goddard}. We place the solitonic core region at the
origin. Let $G$ and $H$ be compact and connected gauge groups, where the
group $H$ is assumed to be embedded in $G$. The gauge group of the core
region $G$ is spontaneously broken to $H$ outside of the core region when
the Higgs field is covariantly constant, $D\Phi =0$. In regions far from the
core ($r\rightarrow \infty $) where we assume that $D^A\Phi _A=0$, it can be
shown that
\begin{equation}
\label{s.s.b.}H^A=\Phi _AF_A,
\end{equation}
where $F_A\in \Lambda ^2(M_3,E_H)$, a two-form on $M_3$ taking values in the
$H$-Lie algebra bundle, denoted by $E_H$ here \cite{Goddard}. An equivalent
expression to (\ref{s.s.b.}) can be written when $D^B\Phi _B=0$. We adopt
units so that $\Phi ^2=1$ when $r>>1$ and where spontaneous symmetry
breaking has occurred. When $G=U(n)$ and $H=U(1)$, $F_A$ becomes a pure
imaginary two-form on $M_3$. The Bogomol'nyi solitons defined by (\ref{bog'}%
) have an energy coming from (\ref{energy}) topologically fixed by
\begin{equation}
\label{bound}
\begin{array}{c}
{\cal E}=-\int_{M_3}d<\Phi _AH^A>-\int_{M_3}d<\Phi _BK^B> \\
=-\int_{S^2}<\Phi _AH^A>-\int_{S^2}<\Phi _BK^B>
\end{array}
\end{equation}
where $S^2$ is a large sphere surrounding the monopole core. Details of the
solitonic particles in this theory now depend on the extent to which they
satisfy the Bogomol'nyi equations (\ref{bog'}) and are `spontaneously
broken'.

Consider first non-singular particle-like solutions to {\it both}
Bogomol'nyi equations in (\ref{bog'}). Substituting (\ref{s.s.b.}) into (\ref
{bound}) and using the normalisation condition $\Phi ^2=1$ for both Higgs
fields, the energy is fixed by $-\int F_A-\int F_B$. Let us conventionally
interpret $\int F_A/2\pi $ as the magnetic charge ($g$), and $\int F_B/2\pi $
as the electric charge ($q$). We can view $F_A$ and $F_B$ as curvatures on
the line bundles determined by $\Phi _A$ and $\Phi _B$, respectively,
because from (\ref{s.s.b.}) they are the projections of $H_A$ and $K_B$ on
the line bundles. The magnetic and electric charges are thereby the Chern
numbers associated to complex line bundles with curvatures $F_A$ and $F_B$,
respectively, and in this theory are quantized at the classical level. The
stability of the solitonic particle is argued from the topological
interpretation that can be given to (\ref{bound}). When both equations in (%
\ref{bog}) are satisfied the topologically stability is assured if either
the electric or magnetic charge is non-vanishing. To this point, we have
used only half of the Bogomol'nyi equations, $D^A\Phi _A=D^B\Phi _B=0$. From
the projective flatness of the curvatures in the Bogomol'nyi equations (\ref
{bog'}), $H^A=K^B=FI_E$, and (\ref{s.s.b.}) we conclude that $F=\varphi
_AF_A=\varphi _BF_B$ where $\Phi _A=\varphi _AI_E$ and $\Phi _B=\varphi _BI_E
$. From this we find that
\begin{equation}
\label{dyon}\int_{S^2}F_A=\int_{S^2}F_B.
\end{equation}
Therefore non-singular, stable, particle-like solutions to the Bogomol'nyi
equations (\ref{bog'}) are dyons. To obtain electric monopoles there would
appear to be two possibilities, both resulting from a weakening of the
Bogomol'nyi equations (\ref{bog'}).

In the first possibility, one or the other of the non-abelian gauge fields, $%
H^A$ or $K^B$, would not break to $U(1)$ in the far-field, so that
non-abelian gauge field would pass unnoticed through the detector. This
corresponds to the case where, for example, $D^B\Phi _B=0$, $%
H^A=K^B=iFI_E=F_B\Phi _B$, but $D^A\Phi _A\neq 0$. The Bogomol'nyi bound in
this case becomes
\begin{equation}
\label{case1}{\cal E}=-\int_{S^2}F_B<\Phi _A\Phi _B>-\int_{S^2}F_B.
\end{equation}
The second term is $-2\pi $ times the electric charge. Since the projective
flatness reduces the gauge group to $U(n)$, we may use the Killing-Cartan
form for the bundle inner product, $<\ >$. The first integral in (\ref{case1}%
) can be made to vanish when $\Phi _A$ is traceless, because $\Phi
_B=\varphi _BI_E$. An explicit example of an electric monopole would be $A=B$
projectively flat, $D^B\Phi _B=0$, and $\Phi _A$ is any sufficiently
differentiable function taking values in $SU(n)$. The full variational field
equations are seen to be satisfied.

The second possibility would allow both gauge fields to break far from the
core using $D^A\Phi _A=D^B\Phi _B=0$, but the two gauge far-fields $%
H^A=F_A\Phi _A$ and $K^B=F_B\Phi _B$ would be completely decoupled so that $%
F_A$ and $F_B$ and their respective topological charges would be unrelated.
That is, the projective flatness in the Bogomol'nyi equations would be
relaxed. The electric and magnetic charges in this case become, respectively,%
$$
2\pi q=-\int_{S^2}F_B,\qquad 2\pi g=-\int_{S^2}F_A,
$$
with no relation between $q$ and $g$ analogous to (\ref{dyon}).

In this paper we have introduced a topological field theory with a
Bogomol'nyi structure permitting BPS electric, magnetic and dyonic
monopoles. From the general arguments in \cite{olive} the particle spectrum
and mass compare favourably with that of the intermediate vector bosons. In
this theory $W^{\pm }$ have energies of opposite sign, so that one is the
anti-particle of the other. The $Z_0$, presumably, corresponds to the case
where $H^A=K^B=iFI_E$, but $D^A\Phi _A\neq 0$ and $D^B\Phi _B\neq 0$. The
gauge far-fields for the $Z_0$ are non-abelian and pass unnoticed through
the detectors, and have a stabilizing energy given by%
$$
{\cal E}=-\int_{S^2}F<\Phi _A>-\int_{S^2}F<\Phi _B>
$$
We do not yet know whether the classical intermonopole forces share much in
common with those for BPS magnetic monopoles, and we do not know whether the
classical BPS topological monopole dynamics possess quantum mechanical
behaviour as was found for BPS magnetic monopoles in \cite{tra}.

\end{document}